The Proceedings of the 14th Topical Conference on Shock Compression of Condensed Matter (TCSCCM) will be published as a special volume by the American Institute of Physics (AIP). The Editors are M.D. Furnish, M.L. Elert, T. P. Russell and C. T. White.

# SHEAR STRESSES IN SHOCK-COMPRESSED COVALENT SOLIDS

**I.I. Oleynik[1], S.V. Zybin[2], M. L. Elert[3], and C. T. White[4]**

[1]*Department of Physics, University of South Florida, Tampa, FL 33620*
[2] *Materials and Process Simulation Center, California Institute of Technology, Pasadena, CA 91125*
[3]*Chemistry Department, U. S. Naval Academy, Annapolis, MD 21402*
[4]*Naval Research Laboratory, Washington, DC 20375*

**Abstract.** Shear stresses are the driving forces for the creation of both point and extended defects in crystals subjected to high pressures and temperatures. Recently, we observed anomalous elastic materials response in shock-compressed silicon and diamond in the course of our MD simulations and were able to relate this phenomenon to non-monotonic dependence of shear stress on uniaxial compression of the material. Here we report results of combined density functional theory (DFT) and classical interatomic potentials studies of shear stresses in shock compressed covalent solids such as diamond and silicon for three low-index crystallographic directions, <100>, <110>, <111>. We observed a non-monotonic dependence of DFT shear stresses for all three crystallographic directions which indicates that anomalous elastic response of shock compressed material is a real phenomenon and not an artifact of interatomic potentials used in MD simulations.

## INTRODUCTION

Shear stresses in shock compressed solids attracted a considerable interest in the shock wave community. It is believed that they are the driving forces for the creation of both point and extended defects in crystals subjected to high pressures and temperatures. Recently, we have performed MD simulations of shock wave propagation in diamond [1] and silicon [2] and discovered a rich variety of materials response. As shock wave intensity increased, four different regimes of shock wave propagation were observed: (i) pure elastic wave, (ii) shock wave splitting into elastic and plastic waves, (iii) anomalous elastic regime, and (iv) overdriven plastic wave with activated solid-state chemistry. The anomalous elastic response is characterized by the absence of plastic deformations: the material remains uniaxially compressed. In the course of our investigations we found that the effective freezing of plastic deformations was related to non-monotonic behavior of shear stresses upon uniaxial compression of diamond and silicon along particular crystallographic directions.

Our MD simulations were performed using the reactive empirical bond order (REBO) potential for diamond [3] and the environment dependent interatomic potential (EDIP) for silicon [4]. The interatomic potentials are usually fitted to the properties of materials at ambient conditions. It is not clear *a priori* that they will work at very high pressures and temperatures, i.e. at conditions where they have not been validated yet. Therefore, it is unclear whether this anomalous elastic response is an artifact of interatomic potentials used in MD simulations or it is a real phenomenon that might be observed in experiments.

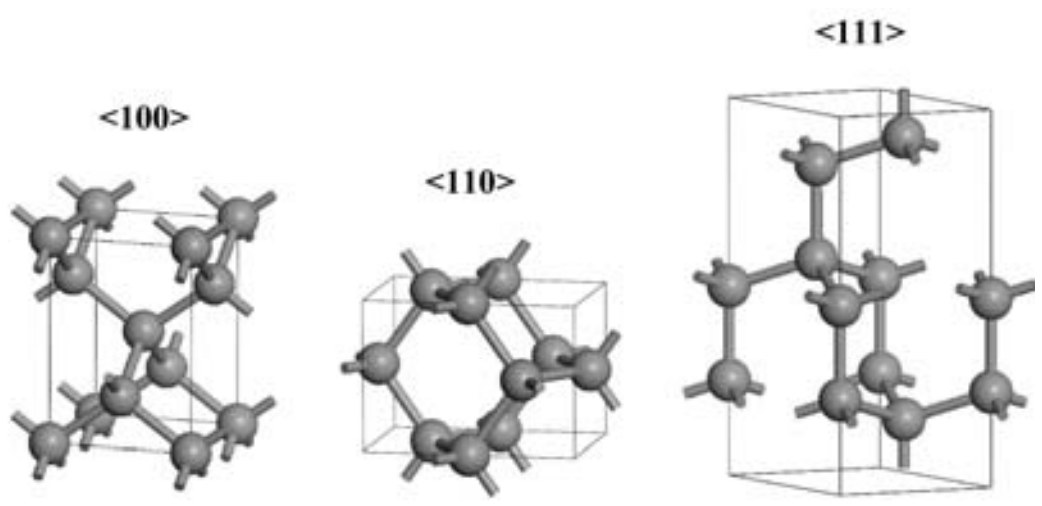

**FIGURE 1.** Unit cells used in calculating the effects of uniaxial compression along the <100>, <110> and <111> directions.

We report results of combined DFT and classical interatomic potential studies of diamond and silicon under uniaxial compression at zero temperature in the <100>, <110> and <111> directions.

## COMPUTATIONAL DETAILS

The shear stresses were calculated by uniaxially compressing diamond and silicon samples in the longitudinal directions while keeping the lateral dimensions fixed. This corresponds to conditions of shock wave experiments at sufficiently large shock wave intensities. The time scale associated with the initial process of shock wave compression is on the order of picoseconds. Therefore, the lattice almost instantaneously transforms to a uniaxially compressed state.

The DFT calculations were performed using the total energy pseudopotential method within a generalized gradient approximation (GGA) density functional [5]. A highly optimized ultrasoft pseudopotential (US-PP) for both silicon and carbon was used with a large plane wave cutoff of 700 eV. Recent work has addressed the question of the accuracy of the pseudopotential approach to calculate properties of matter at extreme conditions. It was found that the US-PP plane wave calculations give almost indistinguishable results from those obtained by all-electron method [6]. The appearance of metallic phases in the course of uniaxial compression requires dense sampling of the k-space Brillouin zone. We used the k-point density 0.02 Å$^{-1}$. The values of energy cutoff and k-space sampling density were chosen to achieve an accuracy of the calculated stresses of better than 0.1 GPa and the energies of better than $10^{-3}$ eV/atom.

The REBO potential [3] was used in classical interatiomic potential simulations of shear stresses in diamond and the EDIP potential [4] for calculation of shear stresses in silicon. Both potentials are considered to be the best classical interatomic potentials currently available for large-scale MD simulations of these materials.

## RESULTS AND DISCUSSION

We calculated the static uniaxial compression of diamond and silicon crystals in the <100>, <110>, and <111> crystallographic directions at zero temperature. The (100), (110), and (111) unit cells used in simulations are shown in Figure 1. The c-axis of a particular unit cell was varied in the appropriate strain interval. For each value of $c$, the lateral dimensions of the cell were kept fixed and all the atomic coordinates were relaxed to have zero net forces on the atoms.

Owing to stress anisotropy, the uniaxial compression in the z-direction creates the shear stresses $\tau_{xz} = 1/2(\sigma_{zz} - \sigma_{xx})$ and $\tau_{yz} = 1/2(\sigma_{zz} - \sigma_{yy})$ that are directed at 45° to the direction of the uniaxial compression. Because of the particular symmetry of the diamond structure, the shear stresses $\tau_{xz}$ and $\tau_{yz}$ are equal to each other for <100> and <110> directions, but different for <111> direction. It is the shear stress that drives the irreversible plastic deformations when exceeding a threshold value.

The DFT and REBO shear stresses for diamond are shown in Figure 2. The DFT maximum and subsequent minimum are at strains 0.4 and 0.6 for the <100> direction, 0.25 and $0.35\,(\tau_{xz})$, $0.4\,(\tau_{yz})$ for the <110> direction, and 0.35 and 0.55 for the <111> direction. The REBO potential shows similar behavior but due to problems with the finite cutoff, the shear profiles are limited to smaller strains.

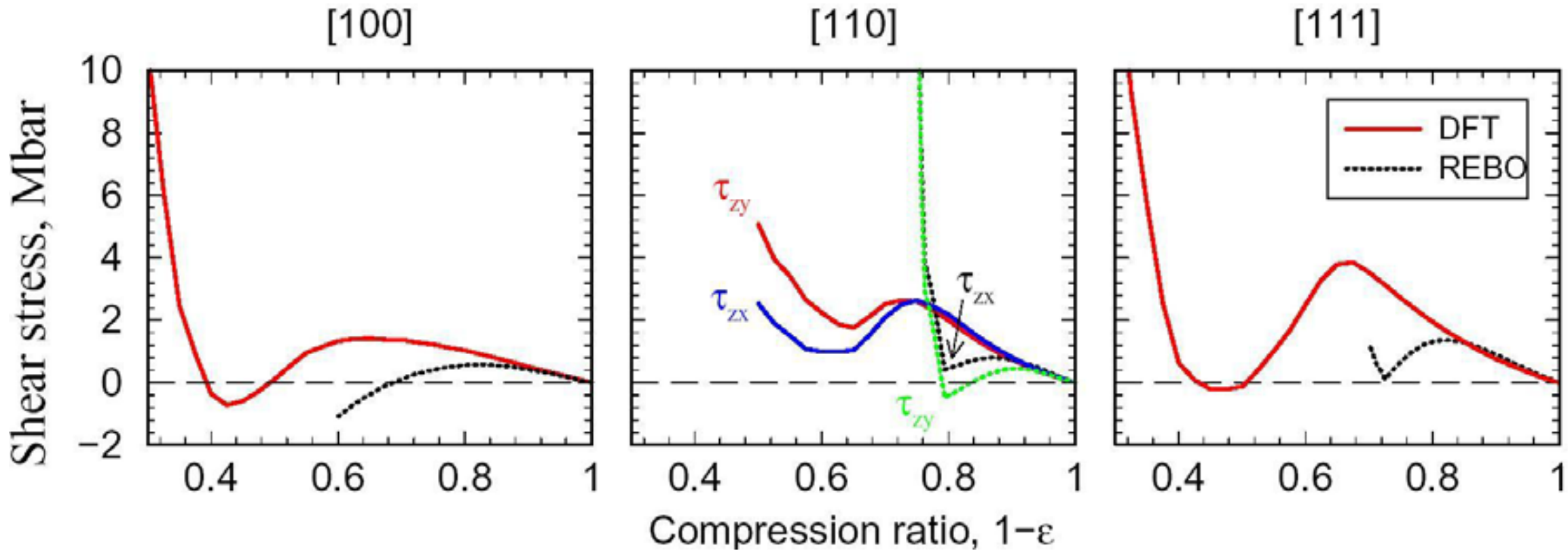

**FIGURE 2.** Shear stresses in diamond for uniaxial compressions along <100>, <110>, and <111> directions.

The silicon results obtained by both DFT and EDIP are shown in Figure 3 together with shear stresses calculated using the popular Stillinger-Weber (SW) silicon potential [7]. We decided to add SW calculations of uniaxial compression of Si because it was used in recent MD simulations of shock wave propagation in silicon [8]. Inspecting Figure 3, one might conclude that the SW potential behaves poorly for lattice strains $\varepsilon > 0.1$ and hence not suitable for simulations of relatively strong shock waves.

The DFT maximum in shear stresses for silicon are at strains $0.175$ for the <100> direction, $0.15$ for the <110> direction, and $0.20$ for the <111> direction. The EDIP shear stresses have similar non-monotonic dependence, but the maxima and minima are observed at different values of strains. The irregular behavior of SW potential and EDIP at large compression ratios is the effect of the short-range cutoff used to determine the number of nearest neighbors for each atom.

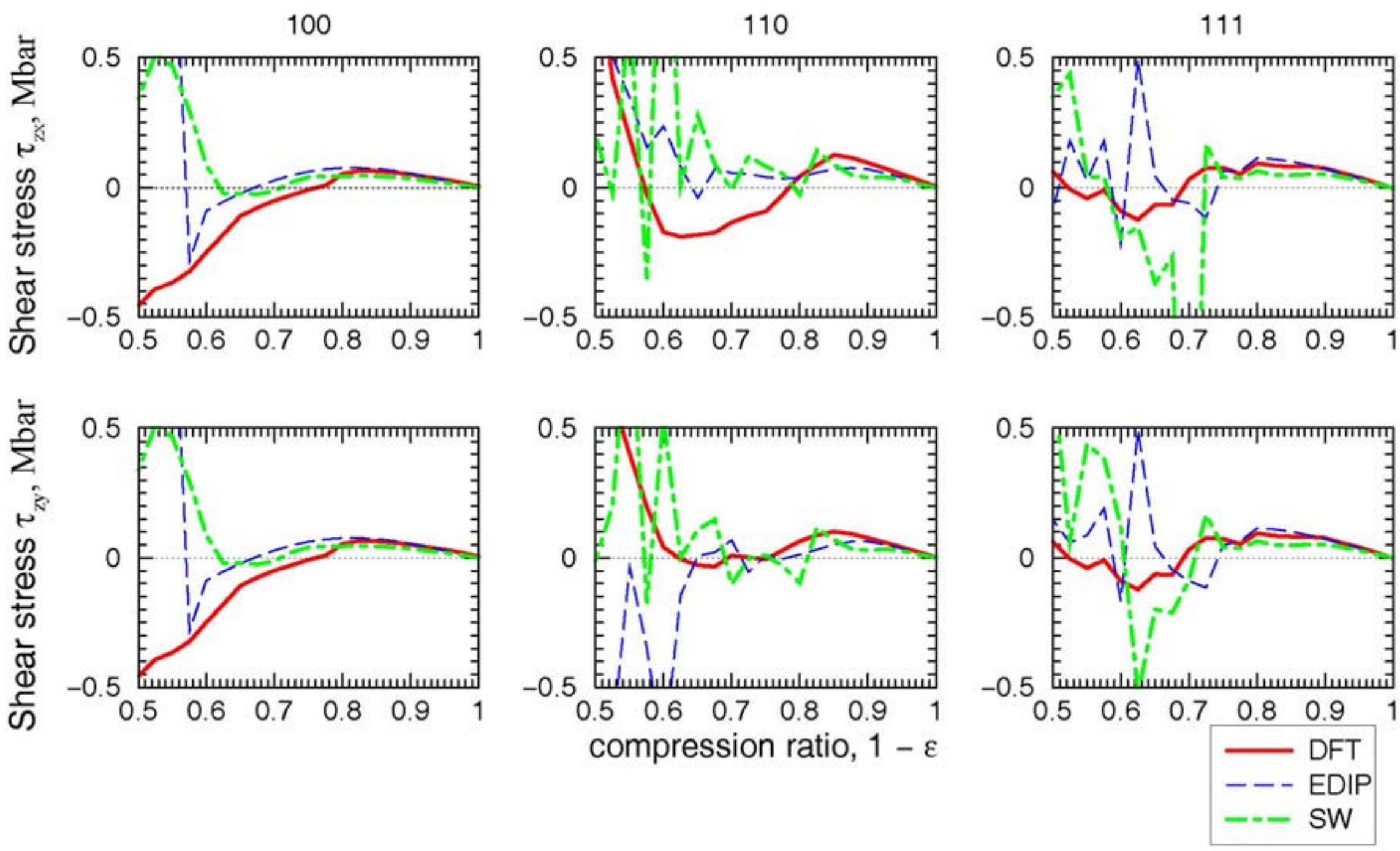

**FIGURE 3.** Shear stresses in silicon for uniaxial compressions along <100>, <110>, and <111> directions.

Importantly, for both DFT and empirical interatomic potentials (REBO or EDIP) the shear stresses pass zero and become negative for the <100> and <111> directions in diamond and for all three directions in silicon. This results in important changes of mechanical properties of compressed diamond. We observed no plastic deformations in our MD simulations exactly in this range of uniaxial compressions. In contrast, in the regime of smaller compressions, deformations occurred by collective displacement of the atoms in the direction of the maximum shear stress. This causes stress relief in the compressed lattice. The same picture is observed at higher uniaxial compressions when the shear stress becomes appreciable again, see Figures 2 and 3.

Because the shear stress is the driving force for the lateral movement and slipping of the crystal planes, the creation of both point and extended defects, there will be some range of uniaxial compressions where these processes are inhibited or even frozen due to very small values of shear stress near the minima of the shear profiles in Fig. 4. We do not exclude the possibility that the zero-shear-stress state of the strained crystal may be metastable, i.e. instability could develop under random thermal movement of the atoms in the shock-heated crystal. We have not observed this development ether in MD simulation of shock propagation or in the quasi-static DFT and REBO (EDIP) relaxation of a uniaxially compressed sample. Evidently, the time scale of this process is outside the time scale accessible by our direct MD simulations. Therefore, special methods are required to address this problem dynamically. In addition, the stability of the uniaxially compressed diamond could also be investigated by evaluating the elastic stability criteria which requires calculation of the phonon spectrum of compressed diamond in addition to the full tensor of elastic constants in each compressed state.

The quantitative comparison of DFT and classical interatomic potentials (REBO for diamond and EDIP for silicon) of small strains show a very good agreement between DFT and REBO (EDIP) as far as energetics and elastic properties are concerned. However, at large strains there are substantial differences, see Figures 2 and 3. Although the REBO and EDIP functional forms include some basic principles of chemical bonding, such as coordination and angular dependence of the bond order terms as well as basic mechanisms of bond breaking and remaking, they are still empirical in nature, that is their parameters were determined by fitting a large database of physical and chemical properties of carbon and silicon systems. The near equilibrium properties of both diamond and silicon are reproduced well because they were included in fitting. However, the properties of C and Si systems at large pressures and temperatures were not taken into account. Therefore, it is not surprising that substantial problems arise at large uniaxial strains. Therefore we conclude that to improve the predictive power of MD shock simulations, further work is required to develop robust and transferable interatomic potentials that are capable of describing systems at high pressures and temperatures.

## ACKNOWLEDGEMENTS

IIO is supported by NSF-NIRT (ECS-0404137) and ARO-MURI (W901 1NF-05-1-0266). Funding at Caltech was provided by ONR and ARO-MURI. CTW is supported by ONR directly and through Naval Research Laboratory.